\documentclass[12pt]{article}

\usepackage{amssymb}

\begin{document}

\begin{titlepage}

\begin{flushright}
\end{flushright}
\vskip 2.5cm

\begin{center}
{\Large \bf Lorentz Violation and $\alpha$-Decay}
\end{center}

\vspace{1ex}

\begin{center}
{\large Brett Altschul\footnote{{\tt baltschu@physics.sc.edu}}}

\vspace{5mm}
{\sl Department of Physics and Astronomy} \\
{\sl University of South Carolina} \\
{\sl Columbia, SC 29208} \\

\end{center}

\vspace{2.5ex}

\medskip

\centerline {\bf Abstract}

\bigskip

Relating the effective Lorentz violation coefficients for composite particles to the
coefficients for their constituent fields is a challenging
problem. We calculate the Lorentz violation coefficients relevant to the
dynamics of an $\alpha$-particle in terms of proton and neutron coefficients.
The $\alpha$-particle coefficients would lead to anisotropies in the $\alpha$-decays
of nuclei, and because the
decay process involves quantum tunneling, the effects of any
Lorentz violations could be exponentially enhanced.

\bigskip

\end{titlepage}

\newpage

\section{Introduction}

At present, there exists a great deal of interest in the possibility that
Lorentz and CPT invariance may not be exact in nature. If the fundamental physical
laws do not respect these symmetries, then we could expect to see evidence of
Lorentz and CPT violations even in the effective theory that governs
conventional low-energy phenomena. If small violations of these symmetries
were discovered, this would
provide a critically important piece of information about the fundamental
structure of physics and a clue to what other new effects we could expect to see.
There is a parameterization of Lorentz and CPT violations in low-energy effective
field theory, known as the standard model
extension (SME), which contains possible Lorentz- and CPT-violating corrections to
the standard model~\cite{ref-kost1,ref-kost2} and general
relativity~\cite{ref-kost12}. Both the
renormalizability~\cite{ref-kost4,ref-colladay2} and
stability~\cite{ref-kost3} of the SME have been carefully examined.

The SME provides a useful framework for interpreting experimental tests of
Lorentz and CPT symmetry.
Sensitive searches for Lorentz violation have included studies of matter-antimatter
asymmetries for trapped charged
particles~\cite{ref-bluhm1,ref-gabirelse,ref-dehmelt1} and bound state
systems~\cite{ref-bluhm3,ref-phillips},
determinations of muon properties~\cite{ref-kost8,ref-hughes}, analyses of
the behavior of spin-polarized matter~\cite{ref-kost9,ref-heckel2},
frequency standard comparisons~\cite{ref-berglund,ref-kost6,ref-bear,ref-wolf},
Michelson-Morley experiments with cryogenic resonators~\cite{ref-antonini,
ref-stanwix,ref-herrmann}, Doppler effect measurements~\cite{ref-saathoff,ref-lane1},
measurements of neutral mesons~\cite{ref-kost10,ref-kost7,ref-hsiung,ref-abe,
ref-link,ref-aubert}, polarization measurements on the light from distant
galaxies~\cite{ref-carroll2,ref-kost11,ref-kost21,ref-kost22},
high-energy astrophysical
tests~\cite{ref-stecker,ref-jacobson1,ref-altschul6,ref-altschul7} and others.
The results of these experiments set bounds on various SME coefficients. Up-to-date
information about bounds on the SME coefficients may be found in~\cite{ref-tables};
at the present time, many of the SME coefficients are quite
strongly constrained, but many others are not.

There are many systems and reaction processes that could potentially be used to
set further bounds on the SME's coefficients for Lorentz violation.
We shall consider a particular system---the helium nucleus, or
$\alpha$-particle---because of the insights it provides into the general properties
of Lorentz-violating physics.

One challenging problem in the study of Lorentz violation is understanding Lorentz
violation for composite particles. The fundamental fields in the SME are the gauge
fields, leptons, quarks,
and the Higgs. However, bounds on Lorentz violation are usually
formulated in terms of bounds on SME coefficients for hadrons, rather than the more
fundamental quark and gluon coefficients. The reason for this is obvious; hadrons are
the physical excitation of the strongly interacting fields. However, they have a
nontrivial structure, and the relationship between the quark, gluon, and hadron
coefficients is not entirely clear.

There are very precise bounds on many Lorentz violation coefficients for the proton
and neutron. However, most measurements of these coefficients are made not on free
nucleons but on more complicated nuclei. Some model must be used to relate the
possible effects of Lorentz violation in the nuclear system to the coefficients for
individual nucleons. This is a simpler problem than relating hadron coefficients to
the coefficients for
subnuclear partons. It is often reasonable to take a simplified model, such as the
Schmidt model~\cite{ref-schmidt},
which assigns all the angular momentum of nucleus to a single
unpaired nucleon constituent. However, this model obviously entails a great deal of
idealization,
and we would like to understand the nature of Lorentz violation in nuclei and other
composite particles more fully.

In this paper, we shall consider one of the simplest composite particles: the helium
nucleus, containing two protons and two neutrons. The simple closed shell structure
of this nucleus ensures that many spin-dependent coefficients must vanish. Yet there
is still some nontrivial structure which we shall uncover. We shall then examine
the possible impact of Lorentz violation on the $\alpha$-decays of nuclei.

There are several motivations for this work. First, it presents a new calculation of
the effective Lorentz violation coefficients for a composite particle, in terms of
the coefficients for its constituents. Second, it examines how the effects of Lorentz
violation can affect the intrinsically quantum mechanical phenomenon of tunneling.
Finally, it suggests a new method for placing laboratory bounds on a number of
Lorentz violation coefficients for protons and neutrons. Although the likely
constraints are not particularly tight, this method offers a way to constrain
several coefficients that have not previously been bounded by laboratory
experiments.

This paper is organized as follows. Section~\ref{sec-Halpha} discusses the
Lorentz violation coefficients relevant for $\alpha$-particles and how these are
related to the coefficients for the constituent protons and neutrons.
Then Section~\ref{sec-rates} shows how these coefficients could impact the
physical process of $\alpha$-decay. In Section~\ref{sec-exp} we look
quantitatively at how $\alpha$-decay studies could be used to place bounds on a
number of Lorentz violation coefficients and in Section~\ref{sec-concl} present
our conclusions.

\section{Lorentz Violation for $\alpha$-Particles}
\label{sec-Halpha}

In this section, we shall derive the effective Hamiltonian for an $\alpha$-particle,
including the effects of Lorentz symmetry violations in the proton and neutron
sectors.
In both this section and the next,
some approximations will be necessary in order to calculate the effects of the
Lorentz violation on $\alpha$-decays.
However, the results will be at least semi-quantitative and good enough to place
order of magnitude bounds.

The minimal SME Lagrange density of a single species of fermion is
\begin{equation}
\label{eq-Lf}
{\cal L}_{f}=\bar{\psi}(i\Gamma^{\mu}\partial_{\mu}-M)\psi,
\end{equation}
where
\begin{equation}
M=m+\!\not\!a-\!\not\!b\gamma_{5}+\frac{1}{2}H^{\mu\nu}\sigma_{\mu\nu}+im_{5}
\gamma_{5},
\end{equation}
and
\begin{equation}
\Gamma^{\mu}=\gamma^{\mu}+c^{\nu\mu}\gamma_{\nu}-d^{\nu\mu}\gamma_{\nu}
\gamma_{5}+e^{\mu}+if^{\mu}\gamma_{5}+\frac{1}{2}g^{\lambda\nu\mu}
\sigma_{\lambda\nu}.
\end{equation}
Electromagnetic interactions are introduced via the minimal coupling substitution
$p^{\mu}\rightarrow p^{\mu}-qA^{\mu}$, where $q$ is the charge.
A nonrelativistic Hamiltonian $H_{f}$ may be derived from (\ref{eq-Lf}) using a
Foldy-Wouthuysen transformation~\cite{ref-foldy}; this effective Hamiltonian
is~\cite{ref-kost26}
\begin{eqnarray}
H_{f} & = &\frac{p^{2}}{2 m}+ \left[m\left(-c_{jk} -
\frac{1}{2}c_{00}\delta_{jk}\right)\right]
\frac{p_{j}p_{k}}{m^{2}}
\nonumber \\
& & + \left[\left(-b_{j}+ m d_{j0}-\frac{1}{2}m\epsilon_{jkl}
g_{kl0}+ \frac{1}{2} \epsilon_{jkl} H_{kl}\right)\sigma^{j}
-a_{j}- m\left(c_{0j} + c_{j0}\right) + m e_{j}\right]\frac{p_{j}}{m}
\nonumber \\ 
& & -\left[b_{0}\delta_{jk} - m \left(d_{kj}+d_{00}\delta_{jk}\right) 
- m \epsilon_{klm}\left(\frac{1}{2}g_{mlj} + g_{m00}\delta_{jl}\right)-
\epsilon_{jkl}H_{l0}\right]\frac{p_{j}\sigma^{k}}{m}
\nonumber \\
& & +\left\{\left[m\left(d_{0j} + d_{j0}\right)-\frac{1}{2}\left(b_{j}+md_{j0}
+ \frac{1}{2}m \epsilon_{jmn} g_{mn0}+\frac{1}{2}\epsilon_{jmn}H_{mn}
\right)\right]\delta_{kl} \right.
\nonumber \\
\label{eq-Hf}
& & \left.+\frac{1}{2}\left(b_{l}+\frac{1}{2}m\epsilon_{lmn}g_{mn0}\right)
\delta_{jk}-m\epsilon_{jlm}\left(g_{m0k} + g_{mk0}\right)\right\}\frac{p_{j}
p_{k}\sigma^{l}}{m^{2}}.
\end{eqnarray}
This is the free nonrelativistic SME Hamiltonian for a single proton or neutron, to
leading order in the SME coefficients.

Many of the terms in (\ref{eq-Hf}) will not contribute when two protons and two
neutrons are combined to form an $\alpha$-particle. The effective Hamiltonian for
the composite particle, $H_{\alpha}$, involves a sum of four $H_{f}$
Hamiltonians---one for each of the nucleon constituents---plus additional
contributions due to nuclear binding effects. We shall not consider the binding in
detail; it's role will simply be to ensure that the four nucleons follow essentially
identical spacetime trajectories.
Moreover, to an excellent approximation,
the two protons and two neutrons in this nucleus are each separately in a spin
singlet state.

So the operator to create an $\alpha$-particle at $\vec{x}$ is
$a^{\dag}(\vec{x})=2b^{\dag}_{p\uparrow}(\vec{x})
b^{\dag}_{p\downarrow}(\vec{x})b^{\dag}_{n\uparrow}(\vec{x})
b^{\dag}_{p\downarrow}(\vec{x})$, where $b^{\dag}_{p\uparrow}$
(or $b^{\dag}_{n\downarrow}$) is the creation operator for a spin up proton
(or spin down neutron). Because of the anticommutivity of the fermion
operators,
\begin{equation}
b^{\dag}_{\uparrow}(\vec{x})b^{\dag}_{\downarrow}(\vec{x})=\frac{1}{2}\left.
\left[b^{\dag}_{\uparrow}(\vec{x}_{1})b^{\dag}_{\downarrow}(\vec{x}_{2})-
b^{\dag}_{\uparrow}(\vec{x}_{2})b^{\dag}_{\downarrow}(\vec{x}_{1})\right]\right|
_{\vec{x}_{1}=\vec{x}_{2}=\vec{x}},
\end{equation}
and $a^{\dag}$ produces an excitation with the total proton spin and total neutron
spin both equal to zero.

Each of the two protons carries the same
mechanical momentum and likewise for
the two neutrons. (Since the proton and neutron masses are very similar, each nucleon
carries approximately one-fourth of
the total momentum of the $\alpha$-particle.)
Consequently, when the spin-dependent terms in each
$H_{f}$ are added together, they give contributions to $H_{\alpha}$ which depend only
on $\vec{\sigma}_{p1}+\vec{\sigma}_{p2}$ and
$\vec{\sigma}_{n1}+\vec{\sigma}_{n2}$---the total proton and neutron spin vectors
in the nucleus. But in the spin-0 singlet state, both of these operators vanish
identically. Therefore, none of the spin dependent terms in $H_{f}$ will contribute
to $H_{\alpha}$.

So we may neglect all spin-dependent terms in $H_{f}$. Doing this, we are left with a
Hamiltonian equivalent to
\begin{equation}
H_{f}\simeq \frac{p^{2}}{2 m} -\left(c_{jk}
+\frac{1}{2}c_{00}\delta_{jk}\right)\frac{p_{j}p_{k}}{m}
-\left[a_{j}+ m\left(c_{0j} + c_{j0}\right) - m e_{j}\right]\frac{p_{j}}{m}.
\end{equation}
However, it is known that the $a^{\mu}$ parameters are unobservable, except in
interactions that involve gravitation or flavor changing
interactions~\cite{ref-kost24}.
The field redefinition
\begin{equation}
\psi\rightarrow e^{-ia\cdot x}\psi,\,\bar{\psi}\rightarrow e^{ia\cdot x}
\end{equation}
eliminates $a$ from the Lagrangian (\ref{eq-Lf}) entirely. This is equivalent to a
translation in
momentum space, $p^{\mu}\rightarrow p^{\mu}-a^{\mu}$. In a theory with $a$ only, the
shifted $\vec{p}$
is the correct mechanical momentum $\gamma m\vec{v}$; including $a$ in the
action merely corresponds to a poor choice of canonical momentum. Since
neither gravity nor the weak interaction are involved with $\alpha$-decay, $a$ will
not contribute to any observable quantity in this kind of decay process.
Moreover, any coefficients
that enter in the same manner as $a$ must also prove unobservable; the whole
expression
$\left[a_{j}- m\left(c_{0j} + c_{j0}\right) - m e_{j}\right]\frac{p_{j}}{m}$ cannot
affect $\alpha$-decay physics. Neglecting this term, we are left with the final
fermion Hamiltonian relevant to $\alpha$-decay,
\begin{equation}
H_{f}\simeq \frac{p^{2}}{2 m}-\left[c_{(jk)}
+c_{00}\delta_{jk}\right]\frac{p_{j}p_{k}}{2m},
\end{equation}
where $c_{(jk)}=c_{jk}+c_{kj}$.
Only the $c$ terms in the fermion sector can affect $\alpha$-decays, and the
remaining terms in
the effective Hamiltonian are all separately invariant under C, P, and T.

The $c_{00}$ term is not observable in solely nonrelativistic experiments; it merely
changes the effective value of $m$. However, we shall retain it, and its effects
can be observed by comparing the kinetic energy of a nonrelativistically moving
particle with the mass energy $m$ observed in particle creation or annihilation
processes.

To find a useful effective Hamiltonian for the $\alpha$-particle, we must take the
sum of four $H_{f}$ terms and then perform a canonical transformation. The
transformation will separate the center of mass motion from the relative motions
of the four nucleons, and it is the center of mass motion that determines the
motion of the $\alpha$-particle.

To see how the center of mass Hamiltonian is modified by $c$-type Lorentz
violation for the constituents, it is simpler to first consider the case of a
two-particle bound state. The problem can be further simplified by considering a
situation in which there is Lorentz violation in the Hamiltonian for only one of
the two constituents.  Once this simple example has been worked out, the
generalization to more constituent particles and more sources of Lorentz violation
is straightforward.

For the two-particle example, the Hamiltonian is
\begin{equation}
\label{eq-H2part}
H=\frac{p_{1}^{2}}{2m_{1}}+\frac{p_{2}^{2}}{2m_{2}}-\left[c_{(jk)}
+c_{00}\delta_{jk}\right]\frac{p_{2j}p_{2k}}{2m_{2}}.
\end{equation}
Since the term in brackets in (\ref{eq-H2part}) is symmetric in $(jk)$ and represents
a small correction, we may diagonalize it by a rotation, choosing
spatial coordinates in which the kinetic energy
for particle 2 has no off-diagonal $p_{j}p_{k}$ terms. In these coordinates, the
kinetic energy splits, as it conventionally does, into three pieces, each of the
form
\begin{equation}
\label{eq-Hj}
H_{j}=\frac{p_{1j}^{2}}{2m_{1}}+\left[1-c_{(jj)}-c_{00}
\right]\frac{p_{2j}^{2}}{2m_{2}}.
\end{equation}
In (\ref{eq-Hj})
and for the remainder of this paragraph, $j$ represents a specific coordinate
and is not to be summed over. $H_{j}$ describes one-dimensional dynamics
equivalent to those for two particles of masses $m_{1}$ and
$m'_{2}=m_{2}/\left[1-c_{(jj)}-c_{00}\right]$. The
effective center of mass coordinate is
$R_{j}=\frac{m_{1}r_{1j}+m'_{2}r_{2j}}{m_{1}+m'_{2}}$, and the conjugate momentum
is the total momentum $P_{j}=p_{1j}+p_{2j}$. The center of mass part of the
Hamiltonian then becomes $P_{j}^{2}/2(m_{1}+m'_{2})$.

Including the dynamics in all three directions, the total center of mass Hamiltonian
is therefore
\begin{equation}
\label{eq-Hcm}
H_{CM}=\frac{1}{2(m_{1}+m_{2})}\left\{\delta_{jk}-\frac{m_{2}}{m_{1}+m_{2}}
\left[c_{(jk)}+c_{00}\delta_{jk}\right]\right\}P_{j}P_{k}
\end{equation}
(once again summing over $j$). The results of the previous paragraph dictate this
this formula holds in coordinates for which $c_{(jk)}$ is diagonal; and since
(\ref{eq-Hcm}) is written in a tensor form, it must also
hold in the original, unrotated coordinate system.

The generalization to more Lorentz-violating particles is straightforward. The
center of mass Hamiltonian will contain a mass-weighted sum of the Lorentz violation
coefficients for the constituents. The weights represent the fraction of the momentum
carried by various constituents, and the sum is the effective Lorentz violation
coefficient for the composite particle. For the $\alpha$-particle, we find
\begin{equation}
\label{eq-Haweighted}
H_{\alpha}=\frac{1}{4(m_{p}+m_{n})}\left\{\delta_{jk}-\frac{m_{p}}{m_{p}+m_{n}}
\left[c^{p}_{(jk)}+c^{p}_{00}\delta_{jk}\right]-\frac{m_{n}}{m_{p}+m_{n}}
\left[c^{n}_{(jk)}+c^{n}_{00}\delta_{jk}\right]\right\}P_{j}P_{k}.
\end{equation}
This does not yet include the effects of nuclear binding. As a first approximation,
we may treat the binding energy $-\epsilon=m_{\alpha}-2m_{p}-2m_{n}$ as if it were
divided up equally among the four constituent
hadrons. This means replacing $m_{p}$ and $m_{n}$ in $H_{\alpha}$ with
$m_{p}-\epsilon/4$ and $m_{n}-\epsilon/4$, respectively. If we also neglect the
neutron-proton mass difference, the effective Hamiltonian for a free
$\alpha$-particle becomes
\begin{equation}
H_{\alpha}=\frac{1}{2m_{\alpha}}\left(\delta_{jk}-\frac{1}{2}k^{\alpha}_{jk}\right)
P_{j}P_{k},
\end{equation}
with $k^{\alpha}_{jk}=c^{p}_{jk}+c^{n}_{jk}+(c^{p}_{00}+c^{n}_{00})\delta_{jk}$.

The $k^{\alpha}_{jk}$ coefficients are nonrelativistic versions of the Lorentz
violation coefficients $k_{\mu\nu}$ for a scalar field. In the presence of
$k_{\mu\nu}$, the Lagrange density for a free relativistic scalar is
\begin{equation}
{\cal L}_{\phi}=\frac{1}{2}(\partial^{\mu}
\phi)(\partial_{\mu}\phi)+\frac{1}{2}k_{\mu\nu}(\partial^{\nu}\phi)
(\partial^{\mu}\phi)-\frac{m^{2}}{2}\phi^{2}.
\end{equation}
However, the coefficients $k_{0j}$ or $k_{00}$ are intrinsically relativistic. They
control violations of boost invariance, and they cannot be measured in a purely
nonrelativistic experiment. Yet the relationship between $c^{n}_{jk}$,
$c^{p}_{jk}$, and $k^{\alpha}_{jk}$ must be the same in every Lorentz frame, since
we may always repeat the preceding analysis with an $\alpha$-particle that is moving
nonrelativistically in the desired frame. In order for this to hold, the full
effective Lorentz violation coefficient for the $\alpha$-particle must be
$k^{\alpha}_{\mu\nu}=c^{p}_{\nu\mu}+c^{n}_{\nu\mu}$. (The change in the placement of
$c_{00}$ terms just corresponds to a change in the overall normalization of the
field and has no physical meaning.) For a more general spin-0 bound state, the
coefficient $k_{\mu\nu}$ will be a mass-weighted sum of the constituents'
$c_{\nu\mu}$ coefficients, like the one appearing in (\ref{eq-Haweighted}).

When each nucleon is minimally coupled to the electromagnetic field by the
replacement $\vec{p}\rightarrow\vec{p}-q\vec{A}$ and $H_{f}\rightarrow H_{f}-qA_{0}$,
the total momentum and $\alpha$-particle Hamiltonian transform as
$\vec{P}\rightarrow\vec{P}+2e\vec{A}$ and $H_{\alpha}\rightarrow H_{\alpha}+2eA_{0}$,
where $e=-|e|$ is the charge of the electron. For nonrelativistic motion in the
absence of an external magnetic field, the contribution of the vector potential
$\vec{A}$ may be neglected.
There can also be interactions between the
$\alpha$-particle and a strong interaction potential $V_{s}$; when included, this
gives the final interacting Hamiltonian
\begin{equation}
H_{\alpha}=\frac{1}{2m_{\alpha}}\left(\delta_{jk}-\frac{1}{2}k^{\alpha}_{jk}\right)
P_{j}P_{k}+2eA_{0}(\vec{x})+V_{s}(\vec{x}).
\end{equation}

\section{Calculating $\alpha$-Decay Rates}
\label{sec-rates}

We shall now proceed to a calculation of how Lorentz violation could impact the
$\alpha$-decay of a nucleus.
There are three major places in which Lorentz violation can enter the decay
process.
These can be loosely related to three constituents of the interaction: the attractive
interaction in the parent nucleus, the expelled $\alpha$-particle, and the
electromagnetic interaction between the $\alpha$-particle and the daughter nucleus.

The decay process can be understood as the $\alpha$-particle tunneling from the
deep attractive potential well of the nucleus, through the Coulomb barrier, and
escaping to infinity.
When the $\alpha$-particle energy is small compared with the height of the barrier,
the details of the potential inside the nucleus are fairly unimportant. We shall
therefore neglect the effects of Lorentz violations on the nuclear potential. In
fact, if the Lorentz violations in various sectors of the SME are comparable in
magnitude (as naturalness conditions would suggest), the dominant Lorentz-violating
contribution to the decay rate $\Gamma$ will arise from one particular factor in our
calculation. The exponential part of the barrier penetration factor depends much
more strongly on the Lorentz violation parameters than any other element in the
formula for $\Gamma$; and the stronger dependence of this factor on $k^{\alpha}$ is
most pronounced when the energy of the decay is smallest compared with the height and
width of the Coulomb barrier.

We shall examine the various factors that together compose the decay rate below.
However, we must first discuss the question of Lorentz violation in the
electromagnetic sector, since electromagnetic
Lorentz violation effects do not become negligible
compared with the effects of $k^{\alpha}$ in any kinematical limit. There is one
form of spin-independent Lorentz violation like $c$ or $k^{\alpha}$ for each
particle species in the standard model. Consideration
only of spin-independent forms is justified by a combination of factors. Since the
$\alpha$-particle has no spin angular momentum, it is automatic that any Lorentz
violations in its sector must be spin independent. This straightforward general
argument was confirmed by the calculations in Section~\ref{sec-Halpha}.
In contrast, the electromagnetic
sector of the SME certainly contains terms that depend on photon polarization.
However, these terms are extremely tightly constrained by astrophysical
polarimetry~\cite{ref-kost11,ref-kost21,ref-kost22}.
Any polarization dependence in the phase speed of light will lead to
birefringence---a change in the polarization of light waves as they propagate.
Birefringence with the right energy dependence to indicate Lorentz violation is not
observed even in radiation that has traversed cosmological distances. Therefore it
is reasonable to neglect these terms in the action, leaving behind an electromagnetic
sector that is spin independent.
With only the conventional terms and spin-independent Lorentz violation, the
Lagrange density for the free electromagnetic sector is
\begin{equation}
{\cal L}_{A}=-\frac{1}{4}F^{\mu\nu}F_{\mu\nu}
-\frac{1}{4}\left(k_{F}\right)^{\beta}\, _{\mu\beta\nu}
\left(F^{\rho\mu}F_{\rho}\,^{\nu}+F^{\mu\rho}F^{\nu}\, _{\rho}\right).
\end{equation}
%

The relevant type of Lorentz violation in each sector can be parameterized by a
two-index
symmetric tensor. However, only the differences between these tensors are observable
physically. A coordinate transformation can be used to eliminate this kind of Lorentz
violation entirely from one sector, at the cost of changing the coefficients in all
other sectors.
Specifically, once the birefringent terms in the electromagnetic sector are set
to zero, the entire sector may be made conventional by a coordinate
redefinition $x^{\mu}\rightarrow x^{\mu}-\frac{1}{2}
\left(k_{F}\right)^{\beta\mu}\,_{\beta\nu}x^{\nu}$. Under such a redefinition,
the Lorentz violation in the $\alpha$-particle sector transforms
$k^{\alpha}_{\mu\nu}\rightarrow k^{\alpha}_{\mu\nu}-
\left(k_{F}\right)^{\beta}\,_{\mu\beta\nu}$. It is this difference of
coefficients that is ultimately measurable; however, for notational simplicity, we
shall henceforth assume $\left(k_{F}\right)^{\beta}\,_{\mu\beta\nu}=0$.

Now we can discuss the details of the $\alpha$-decay process in the presence of
$k^{\alpha}$.
There are two ways of treating the tunneling process by which the $\alpha$-particle
escapes through the electrostatic potential of the nucleus. The tunneling rate may
be determined using either the properties of exact Coulomb wave functions or the
Wentzel-Kramers-Brillouin-Jeffreys (WKBJ) approximation. In generalizing the analysis
of $\alpha$-decay to cover the possibility of Lorentz violation, we shall opt for the
latter approach.

However, this approach immediately runs into a problem.
In the presence of $k^{\alpha}$,
the modified Schr\"{o}dinger is no longer separable (in either spherical or
parabolic coordinates). In order to use the WKBJ approximation, it is generally
necessary to separate the variables; the WKBJ technique can then be applied to one
variable at a time.
So in order to get around this difficulty, we must introduce
another approximation.

The new approximation is a form of the eikonal approximation. The conventional
eikonal approximation, when applied to scattering problems, proceeds as follows. One
examines the linear path that a particle of a given impact parameter would follow
in the absence of interactions. When the interaction is ``turned on,'' a particle
moving along this line would acquire a phase shift---which is just the time integral
of the scattering potential along the straight trajectory. The scattering amplitude
can then be calculated from this phase shift.

In the case of $\alpha$-decay, a different sort of eikonal approximation is required.
To calculate the barrier penetration factor, we shall treat the ejected
$\alpha$-particle as if it were escaping along a well-defined linear path in the
direction of its ultimate velocity. However, since the region of interest along this
path is part of the classically forbidden region, we shall not be calculating a phase
accrued along the path; instead we shall calculate the exponential suppression of the
wave function along the straight line.
This reduces the three-dimensional problem to a one-dimensional
problem for the motion in the direction $\hat{v}$ of the velocity. The kinetic term
in the $\alpha$-particle Hamiltonian for the motion along this axis is
\begin{equation}
K_{\hat{v}}=\frac{1}{2m_{\alpha}}\left(1-\frac{1}{2}k^{\alpha}_{jk}\hat{v}_{j}
\hat{v}_{k}\right)P^{2},
\end{equation}
where $P$ is now the one-dimensional momentum in the relevant direction. [Actually,
in the presence of the Lorentz violation, the directions of the momentum and
the velocity
generally differ; the velocity is $v_{k}=\frac{1}{m_{\alpha}}(p_{k}-\frac{1}{2}
k^{\alpha}_{jk})p_{j}$. However, the difference between the two directions only
contributes a higher order correction to the penetration factor.] For motion purely
in the $\hat{v}$-direction, the effect of the Lorentz violation is to modify the
inertial mass of the $\alpha$-particle to $m_{\alpha}\left(1+\frac{1}{2}
k^{\alpha}_{jk}\hat{v}_{j}\hat{v}_{k}\right)$.

Note that this use of the eikonal approximation is only straightforward if
the nucleus and the $\alpha$-particle are in a relative S state. However, the barrier
penetration factors for $L=0$ angular momentum states are always
greater than for higher angular momentum
states at the same energy, because for $L>0$ states there is an additional
centrifugal term in the effective potential. Because of this (and because of
the related fact that the
wave function for two particles with relative angular momentum $L>0$ vanishes when the
particles' positions coincide), S states frequently dominate in nuclear interaction
processes, and we shall consider only S-wave $\alpha$-decays in our calculations.

Applying the standard WKBJ technique, the barrier penetration factor is then
\begin{equation}
T=\beta^{2}\exp\left[-2\sqrt{2m_{\alpha}}\left(1+\frac{1}{4}k^{\alpha}_{jk}
\hat{v}_{j}\hat{v}_{k}\right)\int_{r_{N}}^{r_{E}}dr\,\sqrt{2(Z-2)e^{2}/4\pi r-E}
\right],
\end{equation}
where
$r_{N}$ and $r_{E}=(Z-2)e^{2}/2\pi E$ are the classical turning points. $r_{N}$
represents the nuclear radius, at which the attractive nuclear potential binding
the $\alpha$-particle to the nucleus comes into play; for a nucleus containing $A$
nucleons, $r_{N}\approx1.4 A^{1/3}$ fm. The escape radius $r_{E}$ is the
beginning of the
classically allowed region outside the repulsive Coulomb potential. The prefactor
$\beta^{2}$ would be negligible for a sufficiently slowly varying potential; while
the exponential arises from the lowest order term in an expansion in powers of
$\hbar$, $\beta$ comes from a higher order term. However, the higher order correction
becomes relevant in the region where the dominant potential changes rapidly from
being the
repulsive Coulomb to the attractive nuclear interaction; taking it into account,
we find $\beta^{2}=\sqrt{(r_{E}/r_{N})-1}$.

The integral in the exponent gives
\begin{equation}
\int dr\,\sqrt{\frac{1}{r}-\frac{1}{r_{E}}}=
r\sqrt{\frac{1}{r}-\frac{1}{r_{E}}}+
\frac{1}{2}\sqrt{r_{E}}
\tan^{-1}\left[\frac{2r-r_{E}}{2\sqrt{r(r_{E}-r)}}\right].
\end{equation}
When $E$ is small compared with the characteristic potential at the nucleus,
$(Z-2)e^{2}/2\pi r_{N}$, the lower limit of integration becomes relatively
unimportant. This limit is equivalent to having $r_{E}\gg r_{N}$, so that the
width of the barrier region is large compared with the size of the nucleus.
The fact that the dependence on $r_{N}$ in this limit is weak is fortuitous, since
the relevant nuclear size can only be determined approximately (and might itself be
affected by Lorentz violation). In the $r_{E}\gg r_{N}$ regime, it is a reasonable
approximation to set $r_{N}\approx0$.
(However, the finite size of the nucleus is an important effect in real
$\alpha$-decays. Some of the earliest measurements of the size of certain
$\alpha$-emitting nuclei were actually based on analyses of the finite nuclear size
corrections to the decay rate.)

In the energy regime for which the penetration factor is relatively
insensitive to the precise value of $r_{N}$, any
Lorentz violation effects in the strong force that holds the parent nucleus together
are of lesser importance.
The most crucial feature of the penetration factor $T$ is that it depends on
$k^{\alpha}$ as
\begin{equation}
T=\beta^{2}\left(\frac{T_{0}}{\beta^{2}}\right)
^{\left(1+\frac{1}{4}k^{\alpha}_{jk}\hat{v}_{j}\hat{v}_{k}\right)},
\end{equation}
where $\beta^{2}T_{0}$ is the penetration factor in the absence of $k^{\alpha}$
Lorentz
violation. $T$ depends exponentially on $k^{\alpha}$; the smaller the penetration
factor is, the larger the fractional dependence on the Lorentz violation will be.

Taking the $r_{N}\approx0$ limit in the exponent, the penetration factor becomes
\begin{eqnarray}
T & = & \beta^{2}\exp\left[-2\sqrt{\frac{m_{\alpha}(Z-2)e^{2}}{\pi}}
\left(1+\frac{1}{4}k^{\alpha}_{jk}\hat{v}_{j}\hat{v}_{k}\right)\left(\frac{\pi}{2}
\sqrt{r_{E}}\right)\right] \\
& = & \beta^{2}\exp\left[-(Z-2)e^{2}\sqrt{\frac{m_{\alpha}}{2E}}
\left(1+\frac{1}{4}k^{\alpha}_{jk}\hat{v}_{j}\hat{v}_{k}\right)
\right],
\end{eqnarray}
with the well-known dependence on $\exp(-bE^{-1/2})$.
The leading corrections to the exponential for finite $r_{N}$ are
\begin{equation}
T=\beta^{2}\exp\left\{-bE^{-1/2}\left[1-\frac{4}{\pi}\left(\frac{r_{N}}{r_{E}}\right)
^{1/2}+\frac{2}{3\pi}\left(\frac{r_{N}}{r_{E}}\right)^{3/2}\right]\right\}.
\end{equation}

The full decay rate is $\Gamma=\frac{\omega}{2\pi}T$, where $\frac{\omega}{2\pi}$ is
the frequency for oscillations of the $\alpha$-particle inside the attractive nuclear
potential region. The frequency gives the rate at which the $\alpha$-particle
strikes the inside of the Coulomb barrier, and $T$ is the probability that it
tunnels through and escapes during a single collision. An estimate of $\omega$ is
$\omega\sim\frac{1}{r_{N}}\sqrt{\frac{E}{m_{\alpha}}}$; this is approximately the
inverse of the time which a particle with kinetic energy $E$ and mass $m_{\alpha}$
takes to traverse a distance $r_{N}$.

Neither $\beta$ nor $\omega$ depend exponentially on the Lorentz violation
coefficients.
We therefore expect the dominant contribution of the Lorentz violation to be made
through its effects on the exponential in
$T$. Neglecting the Lorentz violation in the prefactor $\beta^{2}\omega$, the
decay rate may be written in the simplified form
\begin{equation}
\label{eq-Gamma}
\Gamma=\Gamma_{0}\exp\left(-\frac{(Z-2)e^{2}}{4}\sqrt{\frac{m_{\alpha}}{2E}}
k^{\alpha}_{jk}\hat{v}_{j}\hat{v}_{k}\right),
\end{equation}
where $\Gamma_{0}$ is the rate in the absence of Lorentz violation. In evaluating
$\Gamma_{0}$, many of the approximations (such as $r_{N}\approx0$) that were
used in our determination of the Lorentz-violating correction are not necessary.
In fact, $\Gamma_{0}$ does not need to be calculated at all; it can be taken from
experimental data.

\section{Comparison with Experiment}
\label{sec-exp}

What (\ref{eq-Gamma}) predicts is an anisotropy in the emission of the
$\alpha$-particles. The decay rate depends on the emission direction $\hat{v}$.
Presuming the anisotropy is a small correction,
\begin{equation}
\label{eq-Gamma2}
\Gamma=\Gamma_{0}\left(1-\frac{(Z-2)e^{2}}{4}\sqrt{\frac{m_{\alpha}}{2E}}
k^{\alpha}_{jk}\hat{v}_{j}\hat{v}_{k}\right).
\end{equation}
The effects of Lorentz violation are enhanced by the potentially large parameter
\begin{equation}
\label{eq-bound}
\frac{(Z-2)e^{2}}{4}\sqrt{\frac{m_{\alpha}}{2E}}=0.99(Z-2)\left(\frac{E}{1\,
{\rm MeV}}\right)^{-1/2}.
\end{equation}
To detect an anisotropy, the fractional difference between the decay rates
in different directions must be greater than the fractional error in the measured
rate due to random errors. If $N$ counts are collected,
the signal to noise ratio goes as $\sqrt{N}$, and
repeated measurements made on the same isotope can produce bounds on
$k^{\alpha}$ that scale as $N^{-1/2}$.

To measure an anisotropy in $\Gamma$ is, in principle, straightforward. A sample of
decaying material should be placed inside a detector that would identify the
directions of any outgoing $\alpha$-particles. This directional data would be used to
search for evidence of anisotropy.
Of course, in a laboratory located on the Earth, the planet's rotation must be
accounted for in any absolute directional measurement.
The direction of each decay in the laboratory coordinates must be re-expressed in a non-rotating coordinate system. There is a particular sun-centered celestial
equatorial coordinate system of this nature, in which bounds on Lorentz-violating
coefficients are conventionally expressed. The conversion between laboratory and
sun-centered coordinates is obviously dependent on the sidereal time, and so the time
associated with each decay event must be recorded along with its direction.
Details of the sun-centered coordinates are given in~\cite{ref-bluhm4}.

In practice, the kind of experiment described here could be tricky.
For example, it is desirable to have a small physical sample, so that rescattering of
$\alpha$-particles produced by decays well inside the sample could be minimized.
The ideal $\alpha$-emitter for these purposes is one whose half-life is well known,
which has only one major decay mode, and which has nuclear spin $I=0$ and decays
into another $I=0$ nucleus. The spin condition is important for two reasons. First,
having both the parent and daughter nuclei spinless (as well as the
$\alpha$-particle) ensures that the decay proceeds via an S-wave tunneling state.
Second, if the decaying isotope had a nonzero spin, then a slight polarization of the
population of parent nuclei could lead to a small anisotropy in the decay rate,
which could depend on the relative orientations of the net nuclear spin and the
outgoing $\alpha$-particles' trajectories.

A nuclide with the desired properties is $^{222}$Rn~\cite{ref-akovali}. Its
half-life of 3.8235 days is known with a fractional $1\sigma$ error of less than
$10^{-4}$. $^{222}$Rn is an $I^{P}=0^{+}$ state, which decays overwhelmingly to
$^{218}$Po, another $0^{+}$ isotope. The decay energy is $5.5903\pm0.0003$ MeV.

If any anisotropy in the $\alpha$-decay of $^{222}$Rn can be ruled out with the same
$\lesssim10^{-4}$ accuracy with which the total half-life is known,
(\ref{eq-Gamma2}) and (\ref{eq-bound}) dictate that the Lorentz violation coefficients
$k^{\alpha}$ may be bounded at the level $|k^{\alpha}_{jk}|\lesssim2\times10^{-6}$
(although the trace $k^{\alpha}_{jj}$ would not be constrained, since it does
not lead to any anisotropy).
These bounds are not very strong, especially compared with the results of atomic clock
experiments, but they could give new constraints on neutron and proton $c$
coefficients that have not been bounded in the laboratory.

The reason that an experiment like this would be sensitive to new coefficients
is that the $\alpha$-decay
experiment would involve the measurement of actual decay directions. The
extremely precise clock comparison experiments instead measure energy shifts. The
energy of certain transitions can vary as the angle between the laboratory magnetic
field and a background vector field changes. However, not all possible magnetic
field directions are sampled as the Earth rotates, which is why certain coefficients
are not measured. Constraints on the various coefficients which are difficult to
measure in clock tests can come from cosmic ray observations, because cosmic rays
coming from all possible directions can potentially be
observed~\cite{ref-altschul12}. However, these
astrophysical bounds lack the certainty associated with controlled laboratory
measurements.

\section{Conclusion}
\label{sec-concl}

In this paper, we have examined Lorentz violation for composite particles,
specifically $\alpha$-particles. In low-energy physics, Lorentz violation can
enter the in the $\alpha$-particle sector only through the coefficients
$k^{\alpha}_{\mu\nu}$. These coefficients are linear combinations of the proton
and neutron coefficients $c^{p}_{\nu\mu}$ and $c^{n}_{\nu\mu}$. In the limit of
exact isospin symmetry, $k^{\alpha}_{\mu\nu}=c^{p}_{\nu\mu}+c^{n}_{\nu\mu}$.

More generally, the coefficients for the Lorentz-violating, spin-independent
modification of a composite particle's nonrelativistic
kinetic energy are linear combinations of the
analogous coefficients for its constituent particles. The weight given to each
coefficient in this sum is determined by the fraction of the total momentum that
is carried by the corresponding constituent. This is the solution to a simple
instance of the more general problem of determining the Lorentz violation
coefficients for a composite particle in terms of the coefficients for the
constituents.

The example of the $\alpha$-particle is especially straightforward, because it is a
spin singlet composed of four extremely similar particles. It would be desirable to
get an equivalent understanding of more complicated nuclei. This would be
particularly useful for improving the bounds on Lorentz violation that come from
clock comparison experiments. Such bounds are currently evaluated using
extremely crude models of the nucleus. The scope and accuracy of the resulting bounds
would be significantly improved with a better
understanding of the relationship between the SME coefficients for nucleons and the
hyperfine transition frequencies that can actually be observed in the laboratory.

Even trickier than the problem of relating the effective Lorentz violation
coefficients for nuclei to the coefficients for their hadronic constituents is that
of relating the proton, neutron, and other hadron coefficients to the coefficients for
the underlying quark and gluon fields. These constituents interact strongly through
quantum chromodynamics, and the interaction cannot be treated as a small correction.
Renormalization group effects are also very important; the ultimate relationship
between the Lorentz violation coefficients for the physical hadrons and the
coefficients for the parton fields will depend sensitively on the renormalization
scale. The whole problem is quite difficult;
however, the subject is also extremely interesting, and it represents one of the
most important open problems concerning the structure of the SME.

This paper also examined another phenomenon in quantum mechanics whose interaction
with Lorentz violation has been very little studied: tunneling. In reactions, such
as $\alpha$-decay, whose rate is principally determined by a quantum barrier
penetration factor $T$, the sensitivity to Lorentz-violating effects may be
enhanced. This is a consequence of the penetration factor's exponential
dependence on various parameters. If Lorentz violation leads to a small increase in
the height of the potential barrier through which a particle must tunnel, the
tunneling rate will be diminished by an exponentially greater amount.

While an experiment searching for a possible anisotropy in the emission of
$\alpha$-particles by a decaying isotope is interesting, the bounds that such an
experiment might place on physical Lorentz violation coefficients are not very
precise. Such an experiment would be sensitive to certain coefficients that have not
previously been constrained by laboratory experiments; however, the coefficients in
question have been bounded at the $5\times10^{-14}$ level by observations of
ultra-high-energy cosmic rays.
Constraints based on controlled laboratory
experiments are generally preferable to bounds based on inferences drawn from
astrophysical data, but it is unfortunate that the laboratory tests proposed here
might only give bounds at the $2\times10^{-6}$ level.

However, this work opens the door to the possibility of constraining Lorentz
violation with
other experiments involving tunneling. The ideal experiment for placing
such constraints would have a high barrier, which would make the tunneling rate
more sensitive to any Lorentz violations. To compensate for the low tunneling rate,
a large flux of particles against the barrier would be required. If a process with
the desired characteristics is found, it could be used to place new bounds on
coefficients for Lorentz violation.


\end{document}